\documentclass[aps,prd,amsmath,floats,floatfix,twocolumn,eqsecnum,superscriptaddress,nofootinbib,showpacs]{revtex4}

\usepackage{amssymb}
\usepackage{amsmath}
\usepackage{latexsym}
\usepackage{epsfig}
\usepackage{epstopdf}
\usepackage{longtable}

\begin{document}



\newcommand{\tg}{\tilde{\gamma}}
\newcommand{\tG}{\tilde{\Gamma}}
\newcommand{\tA}{\tilde{A}}
\newcommand{\tR}{\tilde{R}}
\newcommand{\tnabla}{\tilde{\nabla}}

\newcommand{\fg}{\mathring{\gamma}}
\newcommand{\fG}{\mathring{\Gamma}}


\title{Constraint preserving boundary conditions for the
  Baumgarte-Shapiro-Shibata-Nakamura formulation in spherical
  symmetry}

\author{Miguel Alcubierre}
\email{malcubi@nucleares.unam.mx}

\affiliation{Instituto de Ciencias Nucleares, Universidad Nacional
Aut\'onoma de M\'exico, A.P. 70-543, M\'exico D.F. 04510, M\'exico.}

\author{Jose M. Torres}
\email{jose.torres@nucleares.unam.mx}

\affiliation{Instituto de Ciencias Nucleares, Universidad Nacional
Aut\'onoma de M\'exico, A.P. 70-543, M\'exico D.F. 04510, M\'exico.}


\date{\today}


\begin{abstract}
  We introduce a set of constraint preserving boundary conditions for
  the Baumgarte-Shapiro-Shibata-Nakamura (BSSN) formulation of the
  Einstein evolution equations in spherical symmetry, based on its
  hyperbolic structure. While the outgoing eigenfields are left to
  propagate freely off the numerical grid, boundary conditions are set
  to enforce that the incoming eigenfields don't introduce spurious
  reflections and, more importantly, that there are no fields
  introduced at the boundary that violate the constraint equations. In
  order to do this we adopt two different approaches to set boundary
  conditions for the extrinsic curvature, by expressing either the
  radial or the time derivative of its associated ingoing eigenfield
  in terms of the constraints. We find that these boundary conditions
  are very robust in practice, allowing us to perform long lasting
  evolutions that remain accurate and stable, and that converge to a
  solution that satisfies the constraints all the way to the boundary.
\end{abstract}




\maketitle


\section{Introduction}
\label{sec:introduction}

The numerical studies of solutions of the field equations of General
Relativity greatly improved once the different research groups turned
to modified formulations of the Arnowitt-Deser-Misner (ADM) evolution
equations~\cite{York79} in order to have a well-posed strongly
hyperbolic system. Empirically it has been found that the
Baumgarte-Shapiro-Shibata-Nakamura (BSSN)
formulation~\cite{Baumgarte:1998te,Shibata95} is very robust in
practice, and during the last decade it has been employed in many
studies of black hole collisions, dynamics of compact objects, and its
associated gravitational radiation.  The BSSN formulation has two
major differences when compared to the standard ADM evolution
equations: the first one is a conformal decomposition of the metric
and extrinsic curvature in order to isolate the volume element and the
trace of the extrinsic curvature and treat them as independent
variables; the second one is the promotion of the {\em connection
  functions} $\Gamma^i :=\gamma^{mn}\Gamma^i_{mn}$ to independent
variables, plus the fact that their resulting evolution equations are
then modified by adding multiples of the momentum constraints (which
vanish for physical solutions) in order to improve stability. The
robustness of the formulation can be traced to this last step since it
can be shown that it renders the system strongly hyperbolic.

It has been established since Friedrichs pioneering work on the
structure of the initial value problem (IVP) for different reductions
of the Einstein field equations, that the hyperbolicity properties
play a crucial role for the robustness of the resulting evolution
systems (see \cite{Friedrich81,Friedrich85,Friedrich96}, and for
recent reviews \cite{Friedrich:2000qv,Hilditch:2013sba}).  A general
result for hyperbolic systems of partial differential equations
(PDE's) is that strong hyperbolicity implies well-posedness. An
important fact that has to be stressed is that this result assumes
that the functions that define the system are defined on an unbounded
domain.  On the other hand, practical uses of an IVP scheme require
the inclusion of artificial boundaries, and the way we handle the
boundary conditions may spoil the well-posedness of a strongly
hyperbolic system. For this reason many researchers have studied the
formulation of General Relativity as an initial-boundary value problem
(IBVP)~\cite{Friedrich99,Szilagyi00a,Szilagyi:2002kv,Calabrese:2003as,
  Reula:2004nr,Sarbach:2004rv,Beyer:2004sv,Nagy:2006pr,Babiuc:2006wk,Sarbach:2012pr,Winicour:2012ne}. As
a result of these studies it has been demonstrated that for many
formulations, and corresponding boundary conditions, the IVBP is in fact
well-posed. In particular, recent advances towards this end have been
achieved for the case of the Z4
system~\cite{Bona:2010wn,Hilditch:2012fp}, but very little has been
done for the case of the BSSN formulation~\cite{Nunez:2009wn}.

A separate issue that arises from the boundary conditions applied at
artificial boundaries in the case of General relativity is the fact
that they may generally not be adapted to the constraint equations in
the sense that, even if the IBVP is well-posed, there are violations
of the constraints introduced at the boundary that can propagate
inward throughout the whole computational domain. This issue has also
been considered carefully by studying the propagation of the different
fields at the
boundary~\cite{Calabrese:2001kj,Calabrese:2002xy,Babiuc:2006ik,Kreiss:2006mi,Buchman:2006xf,Kreiss:2007cc,Buchman:2007pj}.
However, up to this date it is not yet clear how to construct general
constraint preserving boundary conditions adapted to the BSSN
evolution system that can be easily implemented in a numerical
evolution code.

In this work we analyze the special case of BSSN in spherical
symmetry~\cite{Brown:2009dd,Alcubierre:2010is}, and based on its
hyperbolic structure we obtain a set of boundary conditions that
preserve the constraints, are very simple to implement, and allow one
to perform long-lasting, robust and convergent simulations.  At the
moment our discussion is restricted to the case of asymptotically flat
spacetimes.  Cosmological spacetimes require a somewhat different
treatment, particularly in the gauge and matter sectors, and we will
consider that case in a future work.

This paper is organized as follows: We first introduce in
Section~\ref{sec:BSSN} the BSSN formulation in spherical symmetry as
presented in~\cite{Alcubierre:2010is}, then in
Section~\ref{sec:hyperbolic} we study the characteristic structure of
the evolution system in order to identify the incoming and outgoing
modes. After that, in Sections~\ref{sec:nonpropagating},
\ref{sec:gauge} and~\ref{sec:constraints} we make use of the incoming
modes to apply consistent boundary conditions, and focus on a
particular mode that propagates inwards from the boundary at the speed
of light, and that can be used to apply boundary conditions adapted to
the constraints. Finally, in Section~\ref{sec:examples} we show two
numerical examples for both a vacuum spacetime and a spacetime whose
matter content is a massless scalar field, where we apply the boundary
conditions introduced before. We conclude in
Section~\ref{sec:conclusions}.


\section{BSSN in spherical symmetry}
\label{sec:BSSN}

We will start with a very brief discussion of the BSSN
formulation~\cite{Shibata95,Baumgarte:1998te} in the particular case
of spherical symmetry.  The BSSN formulation is based on a conformal
decomposition of the spatial metric of the form
\begin{equation}
\tg_{ij} = e^{- 4 \phi} \gamma_{ij} \; ,
\label{eq:gammatildedef}
\end{equation}
In the standard version of this formulation, the conformal factor
$\phi$ is chosen in such a way that the determinant of the conformal
metric is unity $\tg=1$. However, this approach is ill-adapted to
curvilinear coordinates where the determinant of the metric is
generally different from unity even in flat space.  In curvilinear
coordinates it is in fact much better to ask for the determinant of
the conformal metric to reduce to its value in flat space (see {\em
  e.g.}~\cite{Brown:2009dd,Alcubierre:2010is}).  If we denote the flat
background metric in the chosen curvilinear coordinates by $\fg_{ij}$,
we will then ask for $\tg = \fg$.  Moreover, we will ask for this to
remain true for all times, so that $\partial_t \tg = 0$, corresponding
to what Brown calls a ``Lagrangian'' BSSN scheme~\cite{Brown:2009dd}
(for an ``Eulerian'' scheme one asks instead that $\tg$ remains
constant along normal lines, {\em i.e.} as seen by the Eulerian
observers).~\footnote{One can account for both cases by introducing an
  extra parameter $\sigma$ as in Ref.~\cite{Alcubierre:2010is}. For
  this analysis we limit ourselves to the Lagrangian case $\sigma=1$.}

Associated to the conformal decomposition of the spatial metric, the
BSSN formulation treats separately the trace of the extrinsic
curvature $K$ and the conformally rescaled traceless part defined as
\begin{equation}
  \label{eq:Atildedef}
  \tA_{ij}:=e^{-4 \phi} \left(K_{ij}-\frac{1}{3}\gamma_{ij} K\right) \; .
\end{equation}

For the particular case of spherical symmetry, we start by writing the
spatial metric as
\begin{equation}
dl^2 = e^{4 \phi} \left( a(r,t) dr^2 + r^2 b(r,t) \: d \Omega^2 \right) \; ,
\label{eq:spheremetric}
\end{equation}
with $a(r,t)$ and $b(r,t)$ positive metric functions, and $d \Omega^2$
the standard solid angle element $d \Omega^2 = d \theta^2 + \sin^2
\theta d \varphi^2$.  The determinants of the physical and conformal
metrics then take the form:
\begin{equation}
\gamma = a b^2 e^{12 \phi} \left( r^4 \sin^2 \theta \right) \; , \qquad
\tg = a b^2  \left( r^4 \sin^2 \theta \right)  \; .
\end{equation}
On the other hand, the determinant of the flat metric in spherical
coordinates can be easily found by setting $a=b=1$ in the expression
for $\tg$ above:
\begin{equation}
\fg = r^4 \sin^2 \theta \; .
\end{equation}
The condition that $\tg = \fg$ now implies that throughout the
evolution we must have $ab^2 = 1$ (this is true for a Lagrangian
evolution, for an Eulerian evolution $ab^2$ can in fact change in the
case of a non-vanishing shift).

The evolution equations for the different dynamical quantities can be
taken directly from~\cite{Alcubierre:2010is}. Taking $\alpha$ as the
lapse function and $\beta^r$ as the radial component of the shift
vector, we find the following evolution equation for the conformal
factor:
\begin{equation}
\partial_t \phi = \beta^r \partial_r \phi + \frac{1}{6} \tnabla_m \beta^m
- \frac{1}{6} \alpha K \; ,
\label{eq:phidot}
\end{equation}
with $K$ the trace of the extrinsic curvature, and where the conformal
divergence of the shift vector is given by
\begin{equation}
\tnabla_m \beta^m = \partial_r \beta^r
+ \beta^r \left( \frac{\partial_r a}{2a} + \frac{\partial_r b}{b}
+ \frac{2}{r} \right) \; .
\label{eq:divbeta}
\end{equation}
The evolution equations for the conformal metric components are
\begin{eqnarray}
\partial_t a &=& \beta^r \partial_r a + 2 a \partial_r \beta^r
- \frac{2}{3} \: a \: \tnabla_m \beta^m - 2 \alpha a A_a , \qquad
\label{eq:adot} \\
\partial_t b &=& \beta^r \partial_r b + 2 b \: \frac{\beta^r}{r}
- \frac{2}{3} \: b \: \tnabla_m \beta^m - 2 \alpha b A_b \; ,
\label{eq:bdot}
\end{eqnarray}
where we $A_a$ and $A_b$ are defined in terms of the traceless
conformal extrinsic curvature $\tA_{ij}$ as:
\begin{equation}
A_a := \tA^r_r \; , \qquad A_b := \tA^\theta_\theta \; .
\end{equation}
Notice that since the tensor $\tA_{ij}$ is traceless by definition,
the following condition must always hold:
\begin{equation}
A_a + 2 A_b = 0 \; .
\label{eq:traceless}
\end{equation}

For the trace of the extrinsic curvature $K$ we have the following
evolution equation
\begin{eqnarray}
\partial_t K &=& \beta^r \partial_r K - \nabla^2 \alpha
+ \alpha \left( A_a^2 + 2 A_b^2 + \frac{1}{3} \: K^2\right) \nonumber \\
&+& 4 \pi \alpha \left( \rho + S_a + 2 S_b \right) \; , \quad
\label{eq:Kdot}
\end{eqnarray}
with $\rho$ the energy density of matter, and $S_a$ and $S_b$ are the
mixed components of the stress tensor: \mbox{$S_a = S^r_r$} ,
\mbox{$S_b := S^\theta_\theta$}.  Notice also that the Laplacian of
the lapse that appears above is with respect to the physical metric,
and takes the form
\begin{eqnarray}
\nabla^2 \alpha &=& \frac{1}{a e^{4 \phi}} \left[ \rule{0mm}{5mm}
\partial_r^2 \alpha \right. \nonumber \\
&-& \left. \partial_r \alpha \left( \frac{\partial_r a}{2a}
- \frac{\partial_r b}{b}
- 2 \partial_r \phi - \frac{2}{r} \right) \right] . \qquad
\label{eq:lapalpha}
\end{eqnarray}

For the traceless part of the conformal extrinsic curvature we first
notice that in fact we only need an evolution equation for $A_a$,
since the traceless condition implies $A_b = - A_a/2$.  We have
\begin{eqnarray}
\partial_t A_a &=& \beta^r \partial_r A_a - \left( \nabla^r \nabla_r \alpha
- \frac{1}{3} \nabla^2 \alpha \right)
+ \alpha \left( R^r_r - \frac{1}{3} R \right) \nonumber \\
&+& \alpha K A_a - \frac{16}{3} \pi \alpha \left( S_a - S_b \right) \; ,
\label{eq:Adot}
\end{eqnarray}
where now
\begin{equation}
\nabla^r \nabla_r \alpha = \frac{1}{a e^{4 \phi}} \left[ \partial_r^2 \alpha 
- \partial_r \alpha \left( \frac{\partial_r a}{2a}
+ 2 \partial_r \phi \right) \right] \; .
\end{equation}

A crucial part of the BSSN formulation is how we write the components of the
Ricci tensor that appear in Eq.~\eqref{eq:Adot} above.  For BSSN in spherical
symmetry we write
\begin{eqnarray}
R^r_r &=& - \frac{1}{a e^{4 \phi}} \left[ \frac{\partial^2_r a}{2a} 
- a \partial_r \Delta^r  - \frac{3}{4} \left( \frac{\partial_r a}{a} \right)^2 \right.
\nonumber \\
&+& \frac{1}{2} \left( \frac{\partial_r b}{b} \right)^2
- \frac{1}{2} \Delta^r \partial_r a + \frac{\partial_r a}{rb}
+ 2 \lambda \left( 1 + \frac{r \partial_r b}{b} \right)
\nonumber \\
&+& 4 \left. \partial^2_r \phi - 2 \partial_r \phi \left( \frac{\partial_r a}{a}
- \frac{\partial_r b}{b} - \frac{2}{r} \right) \right] ,
\label{eq:sphere-Rrr-reg} \\
R &=& - \frac{1}{a e^{4 \phi}} \left[ \frac{\partial^2_r a}{2a}
+ \frac{\partial^2_r b}{b} - a \partial_r \Delta^r
- \left( \frac{\partial_r a}{a} \right)^2
\right. \nonumber \\
&+& \frac{1}{2} \left( \frac{\partial_r b}{b} \right)^2 + \frac{2}{rb}
\left( 3 - \frac{a}{b} \right) \partial_r b \nonumber \\
&+& 4 \lambda + 8 \left( \partial^2_r \phi + ( \partial_r \phi )^2 \right)
\nonumber \\
&-& \left. 8 \partial_r \phi \left( \frac{\partial_r a}{2a}
- \frac{\partial_r b}{b} - \frac{2}{r} \right) \right] .
\label{eq:sphere-RSCAL-reg}
\end{eqnarray}
where $\lambda:=(1-a/b)/r^2$ is a variable used for the regularization
of the origin (but that here we will use just as a shorthand), and
where we have introduced the ``connection vector'' $\Delta^i$, which
is defined in terms of the Christoffel symbols of the conformal metric
$\tG^i{}_{mn}$ and those of the flat background metric $\fG^i{}_{mn}$
as
\begin{equation}
\Delta^i = \tg^{mn} \left( \tG^i{}_{mn} - \fG^i{}_{mn} \right) \; .
\label{eq:BSSN-Deltadef}
\end{equation}
Being defined as a difference of Christoffel symbols, $\Delta^i$ is in
fact a true vector.  In spherical symmetry the only non-trivial
component is $\Delta^r$, which takes the explicit form
\begin{equation}
\Delta^r = \frac{1}{a} \left( \frac{\partial_r a}{2a}
- \frac{\partial_r b}{b} - 2 r \lambda \right) \; .
\label{eq:sphere-Delta-reg}
\end{equation}

In the BSSN formulation, $\Delta^r$ is now promoted to an independent
variable.  Its evolution equation is obtained from its definition
above, but it is then modified using the momentum constraint to
eliminate the divergence of the traceless extrinsic curvature.  In the
particular case of spherical symmetry one then finds
\begin{eqnarray}
\partial_t \Delta^r &=& \beta^r \partial_r \Delta^r - \Delta^r \partial_r \beta^r
+ \frac{1}{a} \: \partial^2_r \beta^r + \frac{2}{b} \:
\partial_r \left( \frac{\beta^r}{r} \right) \nonumber \\
&+& \frac{1}{3} \left( \frac{1}{a} \: \partial_r ( \tnabla_m \beta^m ) 
+ 2 \Delta^r \tnabla_m \beta^m \right) \nonumber \\
&-& \frac{2}{a} \left( A_a \partial_r \alpha
+ \alpha \partial_r A_a \right) 
+ 2 \alpha \left( A_a \Delta^r - \frac{2r}{b} A_\lambda \right) \nonumber \\
&+& \frac{\alpha \xi}{a} \left[ \partial_r A_a
- \frac{2}{3} \: \partial_r K \right.
+ 6 A_a \partial_r \phi \nonumber \\
&+& \left. r^2 A_\lambda \left( \frac{2}{r}
+ \: \frac{\partial_r b}{b} \right) 
- 8 \pi j_r \right] \; ,
\label{eq:Deltadot}
\end{eqnarray}
with $A_\lambda:=(A_a - A_b)/r^2$ another regularization variable that
we only use here as a shorthand, $j_r$ the (physical) covariant
component of the momentum density, and where $\xi$ is an arbitrary
parameter such that $\xi > 1/2$, with the value $\xi=2$ corresponding
to standard BSSN.

Our final system of evolution equations is then given by
equations~\eqref{eq:phidot}, \eqref{eq:adot}, \eqref{eq:bdot},
\eqref{eq:Kdot}, \eqref{eq:Adot} and \eqref{eq:Deltadot}.

Before finishing this section, it is also convenient to write the
specific form of the Hamiltonian and momentum constraints. One finds
\begin{eqnarray}
H &:=& R - \left( A_a^2 + 2 A_b^2 \right) + \frac{2}{3} \: K^2 
- 16 \pi \rho = 0 \; , \qquad
\label{eq:sphere-ham} \\
M &:=& \partial_r A_a - \frac{2}{3} \: \partial_r K
+ 6 A_a \partial_r \phi \nonumber \\
&+& r^2 A_\lambda \left( \frac{2}{r} + \: \frac{\partial_r b}{b} \right)
- 8 \pi j_r = 0 \; .
\label{eq:sphere-mom}
\end{eqnarray}

One should also remember that, since we have promoted the quantity $\Delta^r$
to an independent variable, we in fact have another constraint corresponding
to the definition of $\Delta^r$, namely
\begin{equation}
C_\Delta := a \Delta^r - \left( \frac{\partial_r a}{2a}
- \frac{\partial_r b}{b} - 2 r \lambda \right) = 0 \; .
\end{equation}


\section{Characteristic structure of the evolution equations}
\label{sec:hyperbolic}

The first step in constructing our boundary conditions is to study the
characteristic structure of the system of evolution equations given
above. We will do this very schematically, the details for the general
case with no spherical symmetry can be found in~\cite{Alcubierre08a}.

Our system of evolution equations is first order in time but
second order in space, so we start by introducing the following
auxiliary quantities
\begin{eqnarray}
q := \partial_r \ln \alpha \; , &\quad&
\chi := \partial_r \phi \; , \\
d_a := \frac{1}{2} \: \partial_r \ln a \; , &\quad&
d_b := \frac{1}{2} \: \partial_r \ln b \; .
\end{eqnarray}
We will also assume that the radial component of the shift vector
$\beta^r$ is an a priori known function of spacetime.  For the
lapse, on the other hand, we will assume a Bona-Masso slicing condition
of the form~\cite{Bona94b}
\begin{equation}
\partial_t \alpha - \beta^r \partial_r \alpha = - \alpha^2 f(\alpha) K \; ,
\label{eq:BonaMasso}
\end{equation}
with $f(\alpha)>0$ an arbitrary positive function of $\alpha$.

With these considerations we find that, up to principal part (higher
derivatives), the system of evolution equations of the previous
Section becomes
\begin{eqnarray}
\partial_0 q &\simeq& - \alpha f \partial_r K \; , \\
\partial_0 \chi &\simeq& - \frac{1}{6} \: \alpha \partial_r K
+ \frac{\beta^r}{6} \: \partial_r \left( d_a + 2d_b \right) \; , \\
\partial_0 d_a &\simeq& - \alpha \partial_r A_a - \frac{\beta^r}{3} \: \partial_r(d_a+2d_b)\; , \\
\partial_0 d_b &\simeq& - \alpha \partial_r A_b - \frac{\beta^r}{3} \: \partial_r(d_a+2d_b)\; , \\
\partial_0 K &\simeq& - \frac{\alpha}{a e^{4 \phi}} \: \partial_r q \; , \\
\partial_0 A_a &\simeq& - \frac{2 \alpha}{3 a e^{4 \phi}} \: \partial_r
\left( q + d_a - d_b + 2 \chi - a \Delta^r \right) \: , \hspace{5mm} \\
\partial_0 \Delta^r &\simeq& - \frac{4 \alpha}{3 a} \: \partial_r K +
\frac{\beta^r}{3a} \: \partial_r(d_a+2d_b)\; ,
\end{eqnarray}
where we have defined $\partial_0 := \partial_t - \beta^r \partial_r$,
and where the symbol $\simeq$ denotes ``equal up to principal part''.
For simplicity, we have also already restricted ourselves to the
standard BSSN choice $\xi = 2$ (the more general case can be easily
considered, but it is not needed for our present purposes). In order
to close the system we should also give the evolution equations of the
lower order terms $a$, $b$, $\alpha$ and $\phi$, but those evolution
equations don't involve spatial derivatives of the dynamical variables
so that they turn out to be trivial up to principal part.

From the structure of the equations above we can immediately see that
there is a mode that does not propagate at all, namely:
\begin{equation}
  \label{eq:eigen.frozen}
\omega^d := d_a + 2 d_b \; .  
\end{equation}
This is not surprising since we started the discussion assuming a
Lagrangian BSSN scheme characterized by $\partial_t \tg=0$, and the
above combination is the precisely dynamical part of $\partial_r \tg$
(if one considers an Eulerian scheme this eigenfield propagates in the
normal direction with speed $-\beta^r$).  This eigenfield is not
usually discussed since for the standard BSSN formulation in
Cartesian-like coordinates one assumes that $\tg=1$, so that its
spatial derivatives vanish. By relaxing this assumption we now find
that this eigenfield must be considered in the analysis, and its
propagation properties will depend on the way in which the conformal
volume element is assumed to evolve.

Proceeding with the analysis, we find two eigenfields that propagate
along the normal lines with speed $-\beta^r$.  They are
\begin{eqnarray}
\omega^q &:=& q - 6 f \chi -f(d_a+2d_b) \; , \\
\omega^\Delta &:=& \Delta^r - 8 \chi / a -(d_a+2d_b)/a \; .
\end{eqnarray}
We can also identify two propagating eigenfields associated with the
slicing condition, they are
\begin{equation}
\omega^\alpha_\pm = e^{2 \phi} \sqrt{af} \: K \pm q \; ,
\label{eq:gaugemodes}
\end{equation}
and they propagate with the characteristic speeds:
\begin{equation}
\lambda^\alpha_\pm = - \beta^r \pm \alpha e^{- 2 \phi} \sqrt{f/a} \; .
\label{eq:gaugespeed}
\end{equation}
Notice that this speed depends on the slicing choice through the
function $f(\alpha)$ (it is a ``gauge'' speed), and can easily be
larger than the speed of light without any physical consequences (this
in fact happens whenever $f>1$).

The last two eigenfields are:
\begin{eqnarray}
\omega^l_\pm &=& e^{2 \phi} \sqrt{a} \left( A_a
- \frac{2}{3} K \right) \nonumber \\
&\pm& \frac{2}{3} \left( d_a - d_b - a \Delta^r + 2 \chi \right) \; ,
\label{eq:physicalmodes}
\end{eqnarray}
and propagate with the speeds:
\begin{equation}
\lambda^l_\pm = - \beta^r \pm \alpha e^{- 2 \phi} \sqrt{1/a} \; ,
\label{eq:lightspeed}
\end{equation}
which are nothing more than the local coordinate speed of light for
outward and inward moving light rays.  Since these last two modes
propagate with the speed of light one could be tempted to think that
they are related to gravitational waves, but they are not.  Remember
that we are in spherical symmetry so there are no gravitational waves.
In fact, these modes propagate at the speed of light only because we
have chosen standard BSSN with $\xi = 2$, any other choice results in
a different propagation speed.  These modes are related to the way in
which the momentum constraint is added to the evolution equation for
$\Delta^r$, so we should expect them to be related to the violation of
the constraints at the boundary.

It is important to mention that for the propagating eigenfields in
equations~\eqref{eq:gaugemodes} and~\eqref{eq:physicalmodes}, the plus
sign corresponds to modes propagating to the right, that is
``outgoing'' modes, while the minus sign corresponds to modes
propagating to the left, i.e. ``incoming'' modes.  One could naively
think that a good boundary condition would be to simply fix the
incoming modes at the boundary to zero, but this is not the case.  As
we will see below in Section~\ref{sec:gauge}, even for the gauge modes
the fact that the outgoing modes behave as spherical waves that fall
off as $1/r$ implies that the incoming mode does not vanish at the
boundary.  For the modes propagating at the speed of
light~\eqref{eq:physicalmodes} the situation is even more complicated,
since, as we will show in Section~\ref{sec:constraints}, the constraints
fix the form of the incoming mode to something highly non-trivial.

A final remark is in order before starting the discussion of boundary
conditions. We evolve numerically the BSSN system formulated as a PDE
system that is first order in time and second order in space, as
discussed in Section~\ref{sec:BSSN}. As such we cannot apply boundary
conditions directly to the eigenfields, but only to the original
dynamical variables.  Still, the knowledge of the characteristic
structure allows us to introduce well behaved and consistent boundary
conditions, as will be dicussed below.


\section{Boundary conditions for the non-propagating eigenfields}
\label{sec:nonpropagating}

From here on we will assume for simplicity that the shift vector
vanishes at the boundary (in modern simulations of isolated systems
this is generally not true, but the shift is still usually very small
at the boundary so that this is not a bad approximation). Under this
assumption the time and normal direction coincide, so we may consider
the case of the eigenfields that propagate along the time and normal
lines in this Section.

We may distinguish two types of eigenfields: the one that propagates
along the time direction, corresponding to the combination
\begin{equation}
\omega^d := d_a + 2 d_b \; ,
\end{equation}
and the ones that propagate along the normal direction, i.e. those
with characteristic speed equal to $-\beta^r$. In our case these
correspond to the following two combinations of $q$, $\chi$,
$\Delta^r$, $d_a$ and $d_b$:
\begin{eqnarray}
\omega^q &:=& q - 6 f \chi -f(d_a+2d_b)\; , \\
\omega^\Delta &:=& \Delta^r - 8 \chi / a -(d_a+2d_b)/a\; .
\end{eqnarray}
While the first one never propagates regardless of the behavior of the
other variables, for a vanishing shift the remaining two fields also
do not propagate, so that one can in principle evolve all of them
directly at the boundary.

Notice that $q$, $\chi$, $d_a$ and $d_b$ are in fact auxiliary
quantities (spatial derivatives of metric components) that are not
evolved directly in the BSSN formulation.  We have found that in
practice one can simply evolve $\alpha$, $\phi$, $a$ and $b$ all the
way to the boundary without applying any special boundary condition to
them.  For the case of a vanishing shift at the boundary, the
evolution equations for these quantities do not involve any spatial
derivatives, so evolving all the way to the boundary is trivial.

This leaves us with the task of applying a boundary condition for
$\Delta^r$.  We have in fact tried two different approaches:

\begin{enumerate}

\item The simplest approach is to just evolve $\Delta^r$ all the way
  to the boundary, using one-sided differences for the radial
  derivatives of $\alpha$ and $K$ that appear in its evolution
  equation~\eqref{eq:Deltadot} (remember that we are taking $\xi=2$).

\item The other possibility is to reconstruct the non-propagating
  field $\omega^\Delta$ at the boundary in the old time level, evolve
  it directly to the new time level, and then solve for $\Delta^r$
  using the fact the we have already updated $a$, $b$ and $\phi$ so
  that we can calculate their radial derivatives in the new time level
  using one-sided differences.

\end{enumerate}

Both these methods allow us to have stable and convergent evolutions,
and in fact seem to give almost identical results.  But the second
method is more complicated to code, so that in practice we prefer the
first.  In summary, in the case of vanishing shift, the
non-propagating eigenfields can be dealt with in a simple way by just
evolving $\alpha$, $\phi$, $a$, $b$ and $\Delta^r$ all the way to
the boundary, using one-sided differences for the spatial
derivatives that appear in the evolution equation for $\Delta^r$.


\section{Boundary conditions for the gauge eigenfields}
\label{sec:gauge}

Next let us consider the gauge eigenfields:
\begin{equation}
\omega^\alpha_\pm = e^{2 \phi} \sqrt{af} \: K \pm q \; ,
\end{equation}
which propagate with the speeds \mbox{$\lambda^\alpha_\pm = \pm
  \alpha e^{- 2 \phi} \sqrt{f/a}$} (again assuming a vanishing shift).
Clearly, one of the eigenfields propagates outward while the other
propagates inward.  In order to respect causality one can then only
give a boundary condition for the incoming eigenfield.  An obvious
choice would be to simply set the incoming field to zero at the
boundary.  This would work well in the case of plane waves in
Cartesian coordinates, but produces noticeable reflections at the
boundary for the case of spherical waves that decay as $1/r$.

We then take a different approach and assume that asymptotically the
lapse function behaves as a spherical wave. This follows from the fact
that when combining equations~\eqref{eq:Kdot} and \eqref{eq:BonaMasso}
the resulting equation for the lapse (assuming a vanishing shift) is a
wave equation of the form
\begin{eqnarray}
\label{eq:alphaddot}
\partial^2_t \alpha - \alpha^2 f \, \nabla^2 \alpha
&=& \alpha^3 f \: \left[ K^2 \left( 2f + \alpha f^\prime - \frac{1}{3} \right) \right.
\nonumber \\
&& \left. - \frac{3}{2}A_a^2 + 4 \pi \left( \rho + S \right) \right] \; .
\end{eqnarray}
In the asymptotic region one expects the right hand side to be
negligible, so we may safely assume that in an asymptotically flat
scenario the lapse behaves for large $r$ as
\begin{equation}
\alpha \simeq 1 + g(r-vt)/r \; ,
\end{equation}
with $g$ some unknown function, and $v=\alpha e^{-2 \phi} \sqrt{f/a}$.
This condition can be imposed as a Sommerfield type condition of the
form:
\begin{equation}
\left( \partial_t + v \: \partial_r \right) \: \left[ r(\alpha-1)\right] \simeq 0 \;,
\end{equation}
which can be seen, by substituting \eqref{eq:BonaMasso}, to be
equivalent to having a non-vanishing incoming eigenfield given by
\begin{equation}
\omega^\alpha_{-} \simeq \left( \alpha - 1 \right) / r \; ,
\end{equation}
quite independently of the form of the function $g$.

So, for our boundary condition we then first update $\alpha$ all the
way to the boundary, find $\omega^\alpha_{-}$ from the above
expression, calculate $q=\partial_r (\ln \alpha)$ at the boundary using
one-sided differences, and finally solve for $K$ from the definition
of $\omega^\alpha_{-}$:
\begin{equation}
K = \frac{\omega^\alpha_{-} + q}{e^{2 \phi} \sqrt{af}} \; .
\end{equation}

The above procedure is consistent since we are only allowed to apply a
boundary condition for the incoming eigenfield, and we are using this
to solve for $K$.  In practice we have found that this boundary
condition is quite robust and stable, and results in very small
reflections at the boundary which become even smaller as the boundary
is pushed further away.

As a final comment we should mention that since this is a pure gauge
sector, any boundary condition we choose for the incoming eigenfield
would in fact be physically consistent.  Still, minimizing boundary
reflections is desirable as we would then remain in essentially the
same gauge if we move the position of the boundaries.


\section{Constraint preserving boundary condition}
\label{sec:constraints}

In the previous Section we discussed how to apply boundary conditions
for the gauge eigenfields that are mathematically consistent in the
sense that they respect causality, which results in the simple
prescription of allowing the outgoing eigenfield to leave the grid
undisturbed, and choosing boundary data only for the incoming
eigenfield.  We also showed how one can choose this incoming data in a
way that is consistent with having spherical outgoing waves, and how
in such a case just setting the incoming fields to zero is not
correct.

One could try to use the same idea for the case of the remaining
eigenfields~\eqref{eq:physicalmodes}.  It turns out, however, that
having boundary data that respects causality is not enough since in
general relativity we also have to make sure that the constraints are
satisfied at the boundary, and in order for this to be true one can
not choose the incoming field freely.  The first thing to notice is
that we can in fact construct both the radial and the time derivatives
of the eigenfields $\omega^l_\pm$ up to principal part as a
combination of constraints.  For example, by starting from
eq.~\eqref{eq:physicalmodes}, taking a radial derivative, and
rearranging terms one can show that:
\begin{equation}
\partial_r \omega^l_\pm = \sqrt{a} e^{2 \phi} M \mp
\left( \frac{a e^{4 \phi} H}{6} + \frac{\partial_r C_\Delta}{2} \right)
+ P_\pm \; ,
\label{eq:radialmethod}
\end{equation}
with $H$, $M$ and $C_\Delta$ the Hamiltonian, momentum and Delta
constraint defined earlier, and where $P_\pm$ is a source term that
involves only first derivatives of the metric components and
undifferentiated components of the extrinsic curvature, whose explicit
form is somewhat long and not particularly illuminating so we will not
write it here (but it is not difficult to find it using an algebraic
computational package). The above relation is not surprising since the
combinations of constraints appearing on the right-hand side
of~\eqref{eq:radialmethod} correspond precisely to the eigenfields of
the constraint subsystem (see Appendix~\ref{sec:constraint.app}).
Imposing now the condition that the incoming constraint eigenfield
should vanish, we can reduce the last equation (taking the minus sign)
to:
\begin{equation}
\partial_r \omega^l_- = P_- \; .
\label{eq:radialmethod2}
\end{equation}

Similarly, one can take instead a time derivative of $\omega^l_\pm$,
use of the evolution equations of the different dynamical quantities,
and rearrange terms to find
\begin{eqnarray}
\partial_t \omega^l_\pm &=& \left(\beta^r \mp \frac{\alpha}{\sqrt{a} e^{2 \phi}} \right)
\left[ \rule{0mm}{5mm} \sqrt{a} e^{2 \phi} M \right. \nonumber \\
&\mp& \left. \left( \frac{a e^{4 \phi} H}{6}
+ \frac{\partial_r C_\Delta}{2} \right) \right]
+ Q_\pm \; ,
\label{eq:timemethod}
\end{eqnarray}
where as before $Q_\pm$ are sources that involve only lower order
terms.  Imposing again that the incoming constraint eigenfield
vanishes we find:
\begin{equation}
\partial_t \omega^l_- = Q_- \; .
\label{eq:timemethod2}
\end{equation}

Notice that equations~\eqref{eq:radialmethod}
and~\eqref{eq:timemethod} can in fact be combined to yield:
\begin{eqnarray}
\partial_t \omega^l_\pm + \left( - \beta^r \pm \frac{\alpha}{\sqrt{a} e^{2 \phi}}
\right) \: \partial_r \omega^l_\pm \hspace{10mm} && \nonumber \\
= Q_\pm + \left( - \beta^r \pm \frac{\alpha}{\sqrt{a} e^{2 \phi}}
\right)  P_\pm \; ,
\end{eqnarray}
which of course just confirms the fact that $\omega^l_{+}$ and
$\omega^l_{-}$ are indeed outgoing and incoming fields respectively.
\vspace{5mm}

The above results suggest two different approaches for imposing a boundary
condition on the incoming field $\omega^l_{-}$:

\begin{enumerate}

\item \underline{Radial derivative method}: Calculate the value of the
  source term $P_{-}$ at the new time level close to the boundary
  (typically one grid point in), using the local evolved values of the
  extrinsic curvature and numerical radial derivatives of the metric
  components (which can easily be calculated since the metric
  components are evolved all the way to the boundary).  Use this value
  of $P_{-}$ to obtain $\partial_r \omega^l_{-}$ close to the
  boundary, and solve for the value of $\omega^l_{-}$ at the boundary
  using equation~\eqref{eq:radialmethod2} and a finite difference
  approximation to the derivative. For example, if the boundary is
  located at the grid point $N$, and we are using a second order
  finite difference approximation, we first calculate
  \mbox{$P_{-}(N-1)$}, and use this to obtain $\omega^l_{-}(N)$ as
  follows
\begin{equation}
\omega^l_{-}(N) = \omega^l_{-}(N-2) + 2 P_{-}(N-1) \: \Delta r \; .
\end{equation}

Finally, once we have $\omega^l_{-}$ at the boundary we can simply use
equation~\eqref{eq:physicalmodes} to solve for the boundary value of
$A_a$, knowing already the boundary values of $K$, $\Delta^r$ and the
different metric components (from which one can calculate their
radial derivatives using one-sided differences).

\item \underline{Time derivative method}: Calculate the value of the
  source term $Q_{-}$ at the boundary at the same time as we calculate
  the time derivatives of all other fields, using one-sided radial
  derivatives when needed.  Use equation~\eqref{eq:timemethod2} to
  evolve $\omega^l_{-}$ at the boundary for one time step at the same
  time as all other dynamical fields. Finally, solve for the boundary
  value of $A_a$ from~\eqref{eq:physicalmodes}.

\end{enumerate}

Both these methods are quite easy to code, though the method based on
the time derivative is conceptually simpler and one could expect it to
give more accurate results.  At first sight, it could seem there is a
lack of consistency since both source terms $P_{-}$ and $Q_{-}$ depend
explicitly on $A_a$, and we are using them to determine precisely the
value of $A_a$ at the boundary.  However, notice that in the space
derivative method we actually just calculate $P_{-}$ at points close
to the boundary and never at the boundary itself, while for the time
derivative method we calculate $Q_{-}$ at the previous time level, so
that in both cases there is no reference to the boundary value of
$A_a$ at the new time level in the calculation of the source terms.

One could also be worried about the fact that we have simply imposed
the condition that the incoming constraint eigenfield is zero, by
assuming that the combinations appearing in
equations~\eqref{eq:radialmethod} and~\eqref{eq:timemethod} (with the
minus sign) are equal to zero. This approach is perfectly consistent
since we allow the outgoing constraint mode to propagate freely
through the boundary.  Moreover, while fixing the incoming constraint
eigenfield to zero may not be the optimal choice, this only introduces
small reflections of the constraints at the boundaries that converge
to zero as the resolution is increased.


\section{Some numerical examples}
\label{sec:examples}

In the previous sections we have discussed how to apply boundary
conditions to the different fields that respect causality and also
guarantee that no violations of the constraints will be introduced at
the boundaries.  Here we will show a few numerical examples of these
boundary conditions at work.

All the simulations presented below were performed with a numerical
code that solves the BSSN system in spherical symmetry, using a method
of lines with a fourth order Runge-Kutta time integrator, fourth order
centered spatial differences for the interior, and fifth order
one-sided differences at the boundary. The origin is staggered, with
symmetry and regularity conditions imposed as described
in~\cite{Alcubierre:2010is}. In particular, we report the results
using the first method proposed in Section~\ref{sec:nonpropagating}
for the non-propagating fields, and omit the other choice since we
find no noticeable difference. Our simulations were performed on a
numerical domain that goes from $r=0$ to $r=25$, using three different
resolutions for convergence analysis, $\Delta r = 0.1,\, 0.0707,\,
0.05$, with corresponding grids of $250$, $354$ and $500$ spatial
points (these values were chosen in order to refine the discretization
interval by a factor of $\sqrt 2$). In each simulation the time step
was chosen such that~\mbox{$\Delta t= \Delta r/2$}.


\subsection{Pure gauge dynamics}
\label{sec:ex_harmonic}

As a first example we will consider the evolution of Minkowski
spacetime with a non-trivial foliation.  For this evolution we will
make use of the so-called harmonic slicing condition, which can be
written in the form
\begin{equation}
  \partial_t \alpha = - \alpha^2 K \; .
  \label{eq:harmonic_slicing}
\end{equation}
Notice that this is a particular case of the Bona-Masso family of
slicing conditions~\eqref{eq:BonaMasso} with $f(\alpha)=1$ (remember
that we are using a vanishing shift vector). One consequence of this
particular choice is that the gauge eigenfields $\omega^\alpha_\pm$
now propagate with the same speed as the modes $\omega^l_\pm$, namely
the coordinate speed of light.

\begin{figure}[ht]
\centering
  \includegraphics[width=8.5cm]{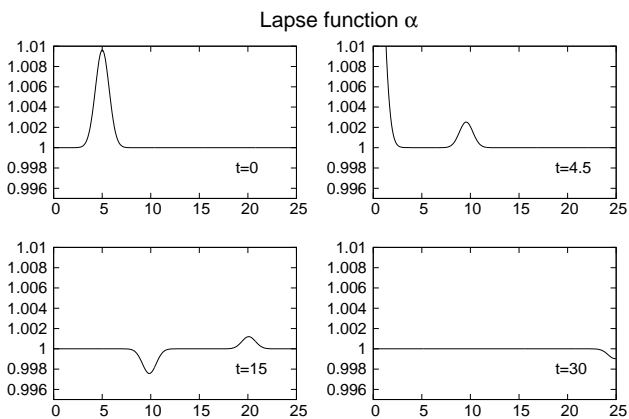}
  \caption{Snapshots of the evolution of the lapse $\alpha$. The
    panels correspond to times $t=0, 4.5, 15, 30$.}
  \label{fig:alpha_harmonic_evol}
\end{figure}

As initial conditions we take a flat slice of Minkowski spacetime such
that
\begin{eqnarray}
  \label{eq:minkowski_initial}
  a = b &=& 1 \;, \nonumber \\
  K = A_a &=& 0 \;, \nonumber \\
  \phi &=& 0 \;, \nonumber \\
  \Delta^r &=& 0 \;, \nonumber
\end{eqnarray}
The only way to obtain a non-trivial evolution is to choose a
non-trivial initial lapse.  We will then take the initial lapse to
have the following Gaussian profile
\begin{equation}
  \label{eq:harmonic_initial}
  \alpha = 1 + \alpha_0\left\{e^{-\left(\frac{r-r_0}{\sigma_0}\right)^2 }
  + e^{-\left(\frac{r+r_0}{\sigma_0}\right)^2 }\right\} \; ,
\end{equation}
where the reflected Gaussian is included in order to have the right
parity behavior of the lapse function at the origin, as demanded by
the symmetry considerations. In this expression $\alpha_0$ is the
amplitude of the Gaussian, $r_0$ its initial position and $\sigma_0$
controls the width.  For the simulation shown below we have used the
parameters $\alpha_0=0.01$, $r_0=5.0$ and $\sigma_0=1.0$; the initial
lapse profile is shown in the first panel of
Figure~\ref{fig:alpha_harmonic_evol}.

We follow the evolution of this initial data in
Figure~\ref{fig:alpha_harmonic_evol}.  Because we are taking $K=0$
initially, the initial Gaussian pulse in the lapse splits into two
smaller pulses moving in opposite directions.  The incoming pulse
grows as it approaches the origin, where it inverts in sign and starts
moving out (panel 2), later both pulses move out with their amplitude
decaying as $1/r$ and eventually leave the numerical domain (panels 3
and 4). The other dynamical fields react accordingly, giving a
non-trivial geometry to the spatial slices at each new time step. The
boundary conditions applied minimize spurious reflections when the
pulses reach the boundary, and the simulation can continue for periods
of time much greater than the light crossing time of the computational
domain, while keeping the simulation accurate and stable.

\begin{figure}[ht]
  \centering
  \includegraphics[width=8.5cm]{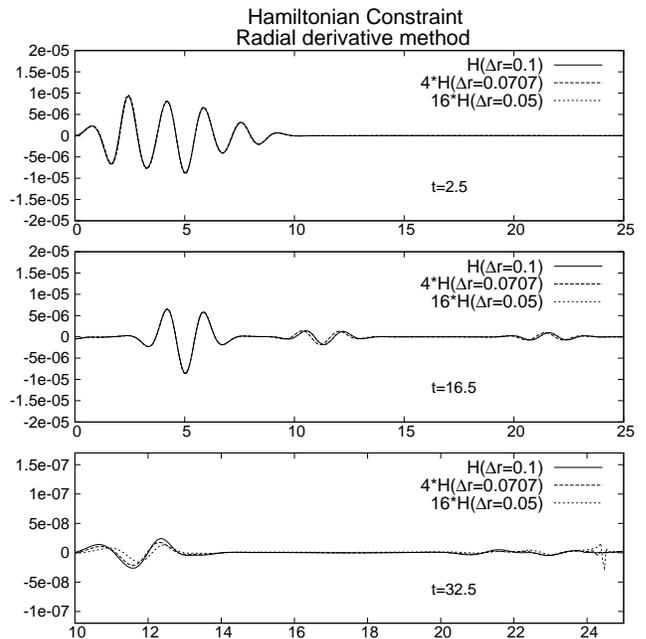}
  \caption{Evolution of the Hamiltonian constraint when applying the
    radial derivative method discussed in the text. The higher
    resolutions have been rescaled by factors of 4 and 16 to show the
    convergence of the solution to fourth order (the ratio between
    subsequent resolutions is $\sqrt 2$). In the last panel the radial
    axis has been shifted and the vertical axis rescaled in order to
    better appreciate the reflected pulses.}
\label{fig:ham_radial_harmonic}
\end{figure}

\begin{figure}[ht]
  \centering
  \includegraphics[width=8.5cm]{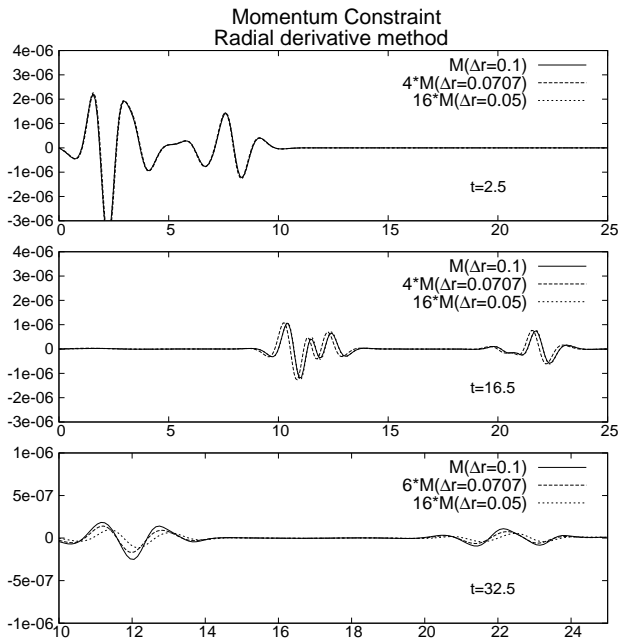}
  \caption{Evolution of the momentum constraint when applying the
    radial derivative method discussed in the text. The lines are
    rescaled as in the plot of the Hamiltonian constraint. In the last
    panel the radial axis has been shifted and the vertical axis
    rescaled in order to better appreciate the reflected pulses.}
\label{fig:mom_radial_harmonic}
\end{figure}

We may further analyze the behavior of the evolution as the
propagating modes reach the outer boundary. First we turn to the
radial derivative method discussed in
Section~\ref{sec:constraints}. Figures~\ref{fig:ham_radial_harmonic}
and~\ref{fig:mom_radial_harmonic} show the numerical evolution of the
Hamiltonian and momentum constraints for the three resolutions
employed.  Of course, since the evolution system propagates the
constraints, the fact that they do not remain equal to zero is only
due to numerical truncation error. As mentioned before, for all our
simulations both the spatial discretization and time integration were
done at fourth order.  This means that when the resolution is doubled,
the values of the constraints should go down by a factor of $2^4=16$.
In the plots we rescale the different resolutions by the corresponding
factors expected for fourth order convergence, so that for a
convergent simulation the different lines should lie on top of each
other.

The first two panels of figures~\ref{fig:ham_radial_harmonic}
and~\ref{fig:mom_radial_harmonic} show the early evolution up to
moments prior to the time when the outgoing pulses reach the
boundary. The convergence of the interior evolution is evident since
the numerical value of the constraints scales consistently with the
fourth order discretization. As mentioned in
Section~\ref{sec:hyperbolic}, in the BSSN formulation there are some
modes that do not propagate. This is made evident on the plots of the
Hamiltonian constraint, where after the initial pulses propagate away
there is always a non-zero remnant at the position of the initial
Gaussian pulse in the lapse (this is essentially caused by the mode
$\omega^d$).  The third panel of both Figures shows the value of the
constraints some time after both outgoing pulses have passed through
the boundary. Notice that in the third panel of both Figures the
radial axis has been shifted, and the vertical axis rescaled, in order
to better observe the reflections induced by the boundaries. The first
thing to notice is that there are indeed reflections of the
constraints at the boundary, and those reflections are much lower in
amplitude (about two orders of magnitude) than the outgoing pulses
that reached the boundary.  Also, the amplitude of those reflections
is smaller as we increase the resolution, and they converge to zero at
the expected fourth order.  We can also appreciate a slight phase
shift in the reflected pulses which is a consequence of the grid
structure employed: since we stagger the origin for regularization
purposes the position of the boundaries does not exactly coincide (for
a given resolution it is shifted $\Delta r/2$ outside of the previous
resolution). We didn't include any higher resolutions since when
doubling the highest resolution shown here the truncation error
becomes comparable with the size of the machine round-off error.

\begin{figure}[t]
  \centering
  \includegraphics[width=8.5cm]{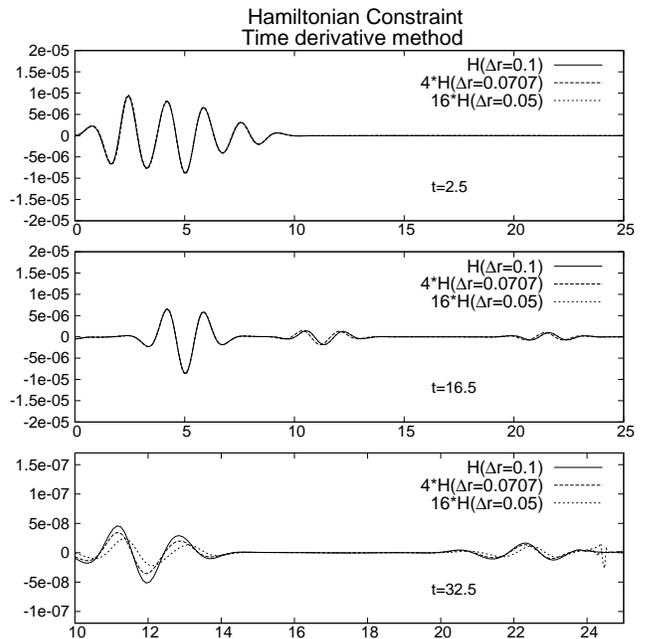}
  \caption{Evolution of the Hamiltonian constraint when applying the
    time derivative method discussed in the text. The higher
    resolutions have been rescaled by factors of 4 and 16 to show
    the convergence of the solution to fourth order (the ratio between
    subsequent resolutions is $\sqrt 2$). In the last panel
    the radial axis has been shifted and the vertical axis rescaled in
    order to better appreciate the reflected pulses.}
\label{fig:ham_time_harmonic}
\end{figure}

\begin{figure}[ht]
  \centering
  \includegraphics[width=8.5cm]{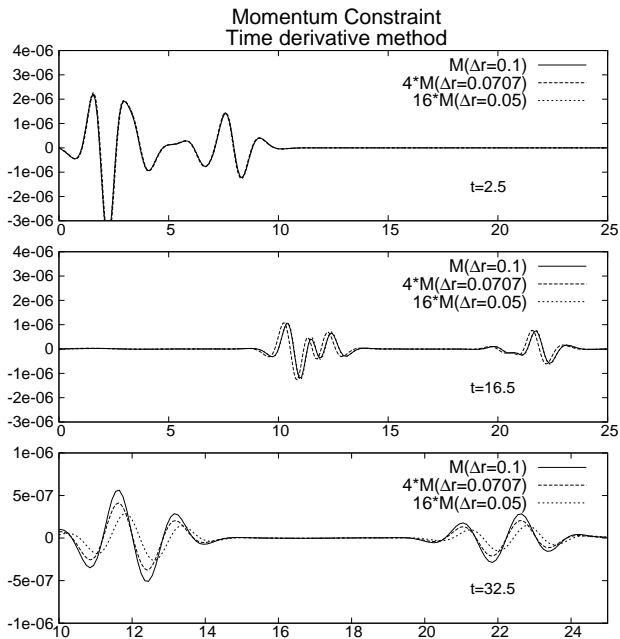}
  \caption{Evolution of the momentum constraint when applying the time
    derivative method discussed in the text. The graphs are rescaled
    as in the plot of the Hamiltonian constraint showing fourth order
    convergence. In the last panel the radial axis has been shifted
    and the vertical axis rescaled in order to better appreciate the
    reflected pulses.}
\label{fig:mom_time_harmonic}
\end{figure}

We now turn to the simulation of the same initial data, but now using
the time derivative method as discussed in
Section~\ref{sec:constraints}. The evolution proceeds in a very
similar way to the previous case, with very small numerical
reflections at the boundaries that converge to zero as the resolution
is increased.  The evolution of the Hamiltonian and momentum
constraints is shown in Figures~\ref{fig:ham_time_harmonic}
and~\ref{fig:mom_time_harmonic} for the three different resolutions
used.  As before, the first two panels of those figures correspond to
times prior to the arrival of the pulses at the boundary, while the
third panel shows the situation some time after the pulses have
reached the boundary. Again, we can see that there are small
reflections. As we increase the resolution, the matching of the
rescaled profiles improves, showing that the violation of the
constraints reflected at the boundaries converges to zero to fourth
order.  The amplitude of the reflections seems to be somewhat larger
than in the previous case (this is more evident in the momentum
constraint), but as the resolution is increased both methods seem to
give very similar results.

\begin{figure}[ht]
  \centering
  \includegraphics[width=8.5cm]{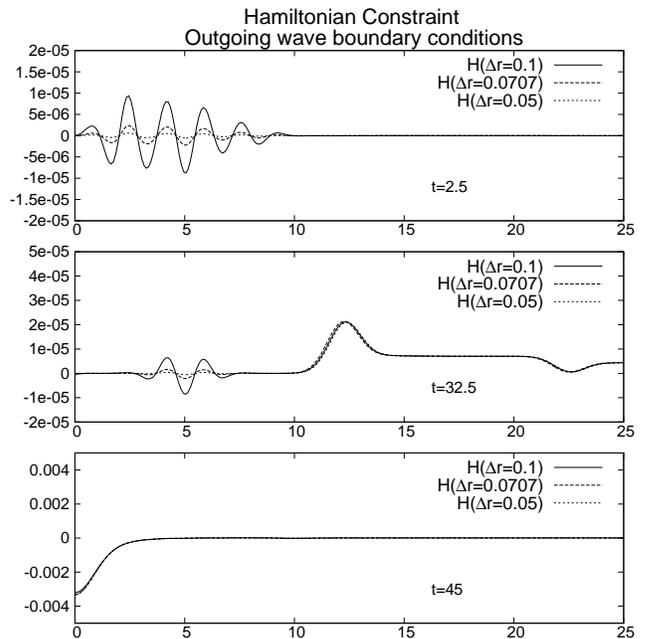}
  \caption{Evolution of the Hamiltonian constraints for a standard
    ``outgoing wave'' boundary condition (this Figure should be
    compared with Figures~\ref{fig:ham_radial_harmonic}
    and~\ref{fig:ham_time_harmonic}). The values are not rescaled as
    in the previous cases in order to show that, once the boundary
    effects become relevant, constraint violations that do not
    converge to zero propagate inward throughout the whole numerical
    domain.}
  \label{fig:Cons.Rad}
\end{figure}

In order to get an idea of how much better our boundary conditions
behave when compared to the more standard ``outgoing wave'' boundary
conditions, we show in Figure~\ref{fig:Cons.Rad} the value of the
Hamiltonian constraint at different times using the following boundary
condition for the different dynamical variables:
\begin{equation}
\left( \partial_t + v \partial_r \right) \left[ r(u-u_0) \right]=0\;,
\end{equation}
where $u$ represents a given dynamical variable, $u_0$ its asymptotic
value, and $v$ the asymptotic characteristic speed (which in this case
is just 1 for all fields). In order to not over-determine the system
at the boundary, the above boundary condition is only applied to the
variables $K$ and $A_a$, while all order variables are evolved up to
the boundary using one-sided differences when necessary. Apart from
the different outer boundary condition, all the other parameters are
identical to the simulations described above. The first panel shows
the situation at $t=2.5$ when boundary effects are not yet relevant.
This should be compared with the first panel of
Figures~\ref{fig:ham_radial_harmonic} and~\ref{fig:ham_time_harmonic},
but notice that now we do not rescale the higher resolutions. We
clearly see that the Hamiltonian converges consistently to zero on the
interior points of the domain.  The situation changes once the
outgoing pulses reach the boundary, as can be seen in the second panel
of Figure~\ref{fig:Cons.Rad} which shows the situation at $t=32.5$,
and should be compared with the third panel of
Figures~\ref{fig:ham_radial_harmonic} and~\ref{fig:ham_time_harmonic}
(but notice the change in scales). From the comparison it is clear
that the reflected constraint violations are much larger than in the
previous cases, by several orders of magnitude, and also that they do
not converge to zero as the resolution is increased.  The last panel
of Figure~\ref{fig:Cons.Rad} shows the situation at \mbox{$t=45$},
once the reflections have reached the origin. Even if the simulation
remains stable and well behaved, the large constraint violations
introduced at the boundary now affect the whole numerical domain and,
since they don't converge to zero, will not allows us to recover
accurately the correct physical solution even in the continuum limit.
It is precisely for this reason that simulations using this standard
boundary conditions must place the boundaries as far away as possible.
We can then conclude that, even for this simple case, our set of
constraint adapted boundary conditions shows a great improvement over
the standard ``outgoing wave'' conditions usually employed in
numerical simulations.

\subsection{Scalar field}
\label{sec:example_scalar}

\subsubsection{Evolution equations and boundary conditions}

For our second example we now consider a truly dynamical scenario that
corresponds to a spacetime whose matter content is a massless scalar
field. The propagation of a spherically symmetric distribution of a
massless scalar field has been well studied in the past, and depending
on the initial data the final stage corresponds to one of two possible
outcomes: for weak data the scalar field disperses away to infinity,
while for stronger data it collapses to form a black
hole~\cite{Choptuik93}. For our purposes it is enough to concentrate
on the non-collapsing case since we are focusing on the interaction at
the outer boundary.

For this case we also need to take into account the effect of the
boundary conditions applied to the scalar field. As before, we choose
a vanishing shift vector. For the lapse function we will now choose
the standard ``1+log'' slicing, which corresponds
to~\eqref{eq:BonaMasso} with $f(\alpha)=2/\alpha$
\begin{equation}
  \partial_t \alpha = - 2 \alpha K \; .
  \label{eq:1+log_slicing}
\end{equation}
The reason for choosing 1+log instead of harmonic slicing is two-fold:
First, 1+log slicing is widely used in numerical relativity so we
believe that it is important to show one example that uses it, and
second, since the asymptotic gauge speed is now $\sqrt{2}$ instead of
1, this example also shows that our boundary conditions behave well
also in a case when not all propagating fields do so at the same
speed.

The matter coupling and the dynamics of the scalar field can be
obtained from its stress-energy tensor
\begin{equation}
T_{\mu \nu} = \partial_\mu \Phi \partial_\nu \Phi
- \frac{1}{2} \: g_{\mu \nu} \: \partial^\alpha \Phi \partial_\alpha \Phi \; .
\end{equation}
The conservation equation ${T^{\mu \nu}}_{; \nu}=0$ implies that the
scalar field obeys the wave equation \mbox{$\Box \Phi = 0$}. Since the
terms that couple to the BSSN equations are quadratic on first
derivatives, the hyperbolic structure of the geometric sector is
unaltered.  In order to rewrite the wave equation for the scalar field
as a first order system in spherical symmetry we first define the
quantities:
\begin{eqnarray}
\Pi &:=& n^\mu \partial_\mu \Phi = \frac{1}{\alpha} \: \partial_t \Phi  \; , \\
\Psi &:=& \partial_r \Phi \; .
\end{eqnarray}
In terms of these quantities, the matter terms that couple to the BSSN
equations are
\begin{eqnarray}
\rho &:=& \frac{1}{2} \left( \Pi^2 + \frac{\Psi^2}{a e^{4 \phi}} \right) \; , \\
j^r &:=& - \Pi \Psi \; , \\
S_a &:=& \frac{1}{2} \left( \Pi^2 + \frac{\Psi^2}{a e^{4 \phi}} \right) \; , \\
S_b &:=& \frac{1}{2} \left( \Pi^2 - \frac{\Psi^2}{a e^{4 \phi}} \right) \; .
\end{eqnarray}

The wave equation rewritten in terms of the first order variables
takes the form:
\begin{eqnarray}
\partial_t \Phi &=& \alpha \Pi \; , \label{eq:scalar_phi_evol}\\
\partial_t \Psi &=& \partial_r \left( \alpha \Pi \right) \; , \label{eq:scalar_psi_evol}\\
\partial_t \Pi &=& \frac{\alpha}{a  e^{4 \phi}}
\left[ \partial_r \Psi + \Psi \left( \frac{2}{r} - \frac{\partial_r a}{2a}
+ \frac{\partial_r b}{b} + 2 \partial_r \phi \right) \right] \nonumber \\
&+& \frac{\Psi}{a  e^{4 \phi}} \: \partial_r \alpha + \alpha K \Pi \; .
\label{eq:scalar_pi_evol}
\end{eqnarray}

Up to principal part the above system just consists on the terms that
contain first derivatives of $\Psi$ and $\Pi$.  We then find that the
eigenfields associated to the scalar field $\omega^\Phi_\pm$, and
their corresponding eigenspeeds $\lambda^\Phi_\pm$, are given by
\begin{equation}
\label{eq:scalar_eigenfields}
\omega^\Phi_\pm = \Pi \mp \frac{\Psi}{e^{2\phi}\sqrt{a}}\,, \quad 
\lambda^\Phi_\pm = \pm \frac{\alpha}{e^{2\phi}\sqrt{a}}\;.
\end{equation}

We will again use these modes to impose our boundary conditions. The
scalar field $\Phi$ is now an auxiliary variable that doesn't
propagate, so integrating its evolution equation all the way up to the
boundary is adequate. Again, just setting the incoming eigenfield
equal to zero is not a good idea since we are in spherical
symmetry. Instead we assume that the scalar field behaves
asymptotically as a spherical wave of the form:
\begin{equation}
\label{eq:phy_asymptotic}
\phi \simeq g(r-v t) / r \; ,
\end{equation}
with $g$ some arbitrary function and $v=\alpha e^{-2
  \phi}/\sqrt{a}$. This can be shown to imply that the incoming field
$\omega^\Phi_-$ behaves as:
\begin{equation}
  \label{eq:scalar_in_boundary}
  \omega^\Phi_- \simeq - \frac{\Phi}{r e^{2\phi}\sqrt{a}} \; .
\end{equation}

With these results we can apply consistent boundary conditions for the
scalar field system. In practice, we integrate the evolution equations
for $\Phi$ and $\Psi$ all the way to the boundary (using one-sided
derivatives as required).  We then we can calculate the incoming
eigenfield at the boundary using
equation~\eqref{eq:scalar_in_boundary} above, and solve for the value
of $\Pi$ at the boundary from the eigenfield
definition~\eqref{eq:scalar_eigenfields}.

\subsubsection{Numerical simulations}

To specify initial data in this case we must first solve the
constraint equations. A simple choice that satisfies the momentum
constraint~\eqref{eq:sphere-mom} is to assume that we start at a
moment of time symmetry, which implies that both the extrinsic
curvature $K_{ij}$ and the time derivative of the scalar field $\Pi$
vanish initially. We are then left with the Hamiltonian
constraint~\eqref{eq:sphere-ham}, which can be further simplified by
choosing the conformal metric to be initially flat. The Hamiltonian
constraint then reduces to a one-dimensional Poisson-type equation for
the conformal factor $\psi \equiv e^{\phi}$:
\begin{equation}
\partial_r^2 \psi + \frac{2}{r} \: \partial_r \psi
+ 2 \pi \psi^5 \rho = 0 \; .
\end{equation}

For a massless scalar field the last equation becomes in fact linear
in $\psi$, since under these assumptions \mbox{$\rho =
  \Psi^2/\psi^4$}. We are now free to specify the initial profile of
the scalar field, which we choose to be a Gaussian distribution
centered on the origin:
\begin{equation}
  \label{eq:scalar_initial}
  \Phi = \Phi_0 e^{-\left( r/\sigma_0 \right)^2} \; .
\end{equation}
We then solve the above equation for the conformal factor assuming
that asymptotically it behaves as $\psi = 1 + c/r$, with $c$ some
constant.

For all the simulations presented below the parameters chosen for the
initial scalar pulse are $\Phi_0 = 0.2$ and
\mbox{$\sigma_0=5.0$}. Since the dynamics are now provided by the
presence of the scalar field, we will also choose the lapse function
to be initially unity, $\alpha(t=0)=1$.  As before, we set up our
numerical grid with boundaries located at \mbox{$r=25$} and use the
same grid parameters as in the previous example. In this case, for
improved stability we add dissipative terms (Kreiss-Oliger
dissipation) when evaluating the right hand side of
equations~\eqref{eq:scalar_phi_evol}, \eqref{eq:scalar_psi_evol}, and
\eqref{eq:scalar_pi_evol}.

\begin{figure}[ht]
  \centering
  \includegraphics[width=8.5cm]{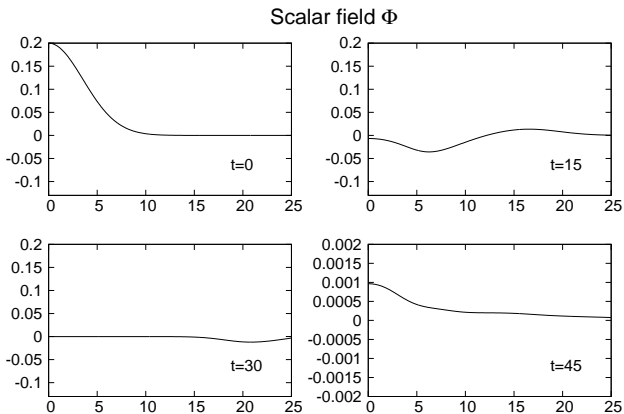}
  \caption{Evolution of the scalar field $\Phi$. When the pulse
    reaches the boundary there is a small reflection, only noticeable
    by rescaling the vertical range on the last panel.}
  \label{fig:scalar_field_evol}  
\end{figure}

\begin{figure}[ht]
  \centering
  \includegraphics[width=8.5cm]{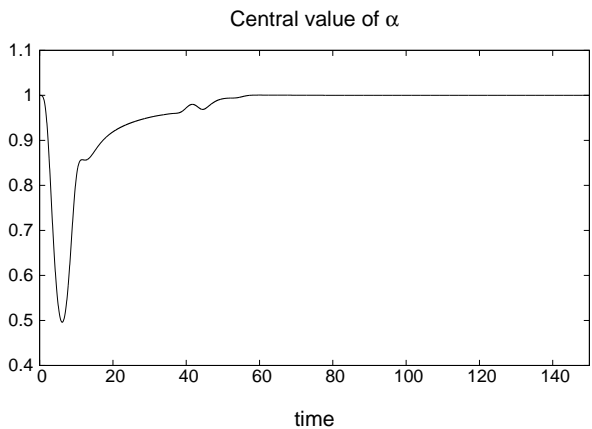}
  \caption{Central value of the lapse function as a function of
    time. Initially the value of the lapse goes down as a consequence
    of the high curvature in the central region, but as the scalar
    field disperses it returns to its original value.}
\label{fig:scalar_alpha0}
\end{figure}

The evolution is somewhat similar to the case of pure gauge dynamics,
the main difference being that it now represents the physical
propagation of matter fields. Figure~\ref{fig:scalar_field_evol} shows
the evolution of the scalar field in the case when we use boundary
conditions based on the radial derivative method. The first panel
shows the initial scalar field gaussian profile.  Panels 2 and 3 show
the situation at $t=15$ and $t=30$, when the outgoing scalar field
pulse is reaching the boundary.  The last panel corresponds to $t=45$,
after the pulse has gone through the boundary. Notice that the
vertical scale has been changed in the last panel in order to show the
small reflections introduced by the numerical boundaries, whose value
is at least two orders of magnitude smaller than the size of the
outgoing pulse as it reached the boundary. We also show on
Figure~\ref{fig:scalar_alpha0} the value of the lapse function at the
origin as a function of time.  As can be clearly seen, the central
value of the lapse initially goes down as a consequence of the
curvature produced by the high matter density and reaches a minimum
value of $0.5$, but the configuration disperses before a trapped
surface can form and the lapse function at the origin returns rapidly
to values close to 1 (this is in fact quite strong initial data, just
slightly short of collapsing to a black hole). Notice that the
outgoing scalar field pulse reaches the boundary at $t \sim 25$, and
the small reflected pulse implodes through the origin at $t \sim 50$.
The effect of these small reflections can barely be seen in the
behavior of the central value of the lapse.

\begin{figure}[ht]
  \centering \includegraphics[width=8.5cm]{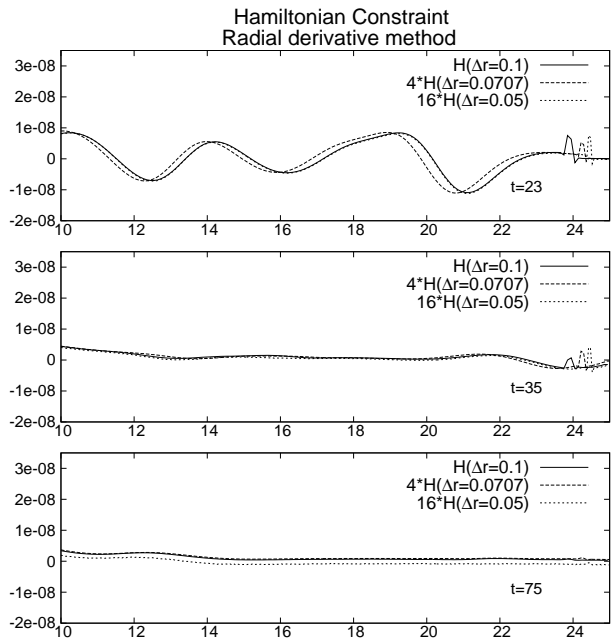}
  \caption{Evolution of the Hamiltonian constraint using the radial
    derivative method. Notice that the radial scale starts at $r=10$.}
  \label{fig:ham_space_scalar}
\end{figure}

\begin{figure}[ht]
  \centering
  \includegraphics[width=8.5cm]{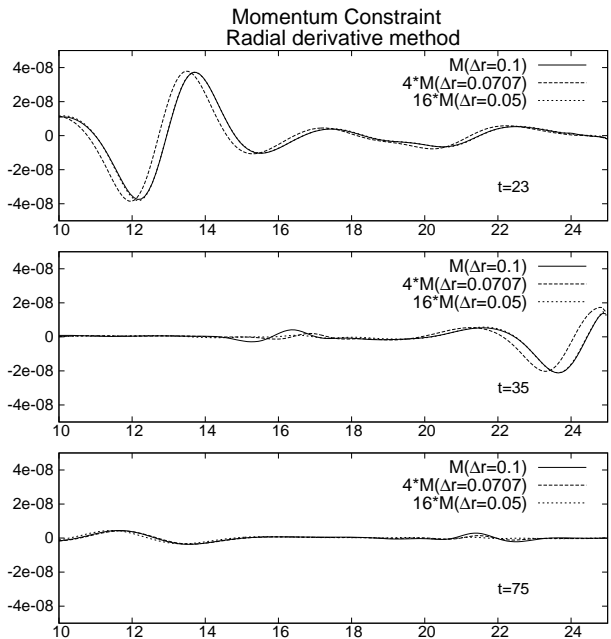}
  \caption{Evolution of the momentum constraint using the radial
    derivative method. Notice that the radial scale starts at $r=10$.}
  \label{fig:mom_space_scalar}
\end{figure}

To further study the behavior of the reflections at the boundaries we
show in Figures~\ref{fig:ham_space_scalar}
and~\ref{fig:mom_space_scalar} snapshots of the evolution of the
Hamiltonian and momentum constraints respectively at times
$t=23,35,75$, for the three different resolutions considered. As
before, the values obtained for each resolution are rescaled by the
appropriate factors to show fourth order convergence.  Notice that in
all three panels of both figures the radial axis starts at $r=10$ in
order to better appreciate the regions closest to the boundary.  The
first panel in both figures shows the situation at $t=23$,
corresponding to a time just before the first pulse reaches the
boundary. We can see good fourth order convergence with the exception
of the two or three points closest to the boundary which appear to be
converging to third order.  The last two panels show the situation at
$t=35$ when the pulses have reflected from the boundary, and at $t=75$
when the reflected pulses have imploded through the origin.

\begin{figure}[ht]
  \centering
  \includegraphics[width=8.5cm]{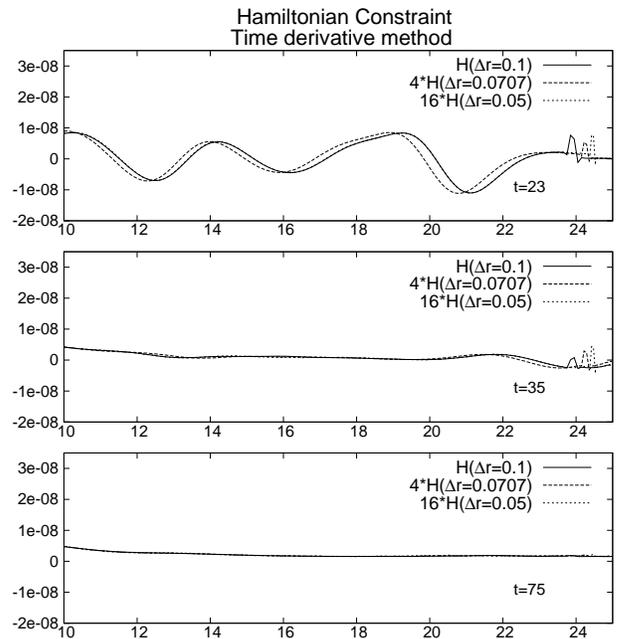}
  \caption{Evolution of the Hamiltonian constraint. Time derivative
    method.}
  \label{fig:ham_time_scalar}
\end{figure}

\begin{figure}[ht]
  \centering
  \includegraphics[width=8.5cm]{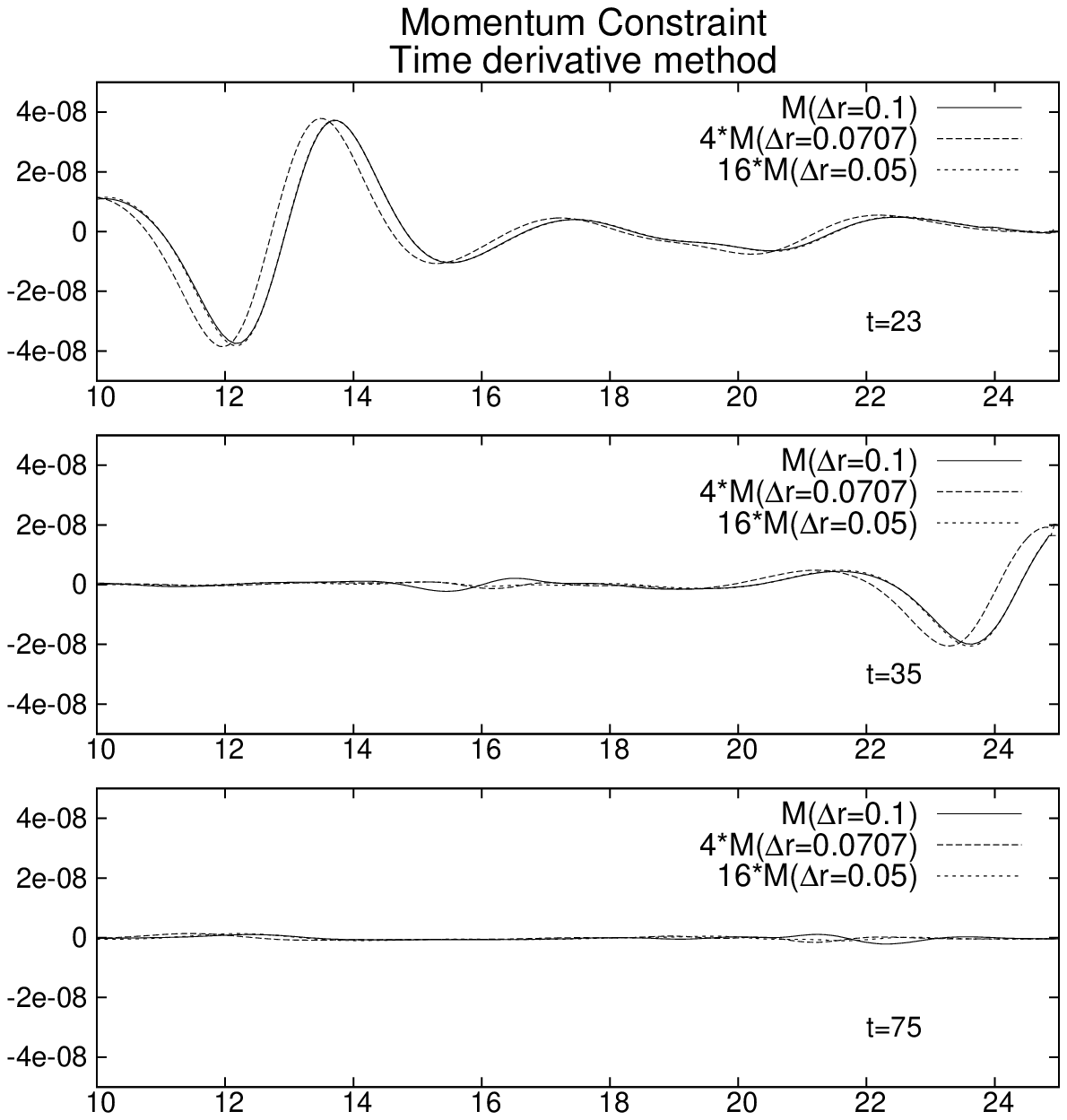}
  \caption{Evolution of the momentum constraint. Time derivative
    method.}
  \label{fig:mom_time_scalar}
\end{figure}

Next, in Figures~\ref{fig:ham_time_scalar}
and~\ref{fig:mom_time_scalar} we consider the same simulations but now
using the boundary conditions based on the time derivative method.
The panels correspond to the same times as before, and the same
rescaling has been used. Again we find good fourth order convergence
at all times.  In particular, the reflections in the momentum
constraint seem smaller than those for the previous case.

The previous examples show that the constraint preserving boundary
conditions work well even when matter fields are present. Although
both schemes work very well in practice, we have found that the time
derivative method shows cleaner convergence, so it may be
preferred over the radial derivative method.


\section{Conclusions}
\label{sec:conclusions}

The aim of this paper was to present boundary conditions that are
derived naturally from the hyperbolic structure of the BSSN evolution
system when restricted to the case of spherical symmetry. The
resulting set of boundary conditions is constraint preserving in the
sense that no spurious constraints violations are introduced at the
boundary.  Even though we do observe that the constraint residuals are
reflected at the boundaries, such reflections converge to zero with
increasing resolution.

The main idea of our method is to first identify the eigenfields of
the evolution system, whose evolution is dictated up to principal part
by simple advection equations.  One can then consider separately the
outgoing, incoming and non-propagating fields at the boundary. The
non-propagating and outgoing fields can be evolved all the way to the
boundary using one-sided derivatives when necessary.  The case of the
incoming fields is more subtle since the naive approach of setting them
equal to zero is only consistent for plane waves in Cartesian
coordinates.  For most applications we may assume that the physical
system studied is isolated, so that the different dynamical fields
behave asymptotically as outgoing spherical waves. This assumption
implies certain conditions for a subset of the incoming fields that
can be used to reconstruct those fields at the boundary.

On the other hand, there exists a special type of incoming fields that
need to be considered separately: those whose radial and time
derivatives at the boundary can be written up to principal part as a
combination of the constraints.  For these fields one is not free to
impose any boundary conditions, since the fact that the constraints
should vanish fixes completely their evolution.  In order to find
those fields at the boundary one first imposes the condition that the
incoming constraint combination should vanish. One can then follow two
different routes: either we reconstruct the incoming fields from their
radial derivative at the boundary, or we integrate them forward in
time using the fact that when the constraints vanish they evolve only
through source terms.

In order to show how these boundary conditions work in practice we
have presented a series of numerical simulations done with our BSSN
code in spherical symmetry, both for a vacuum spacetime with
non-trivial gauge dynamics, and a non-vacuum spacetime with a real
scalar field as the source of the gravitational field. We have shown
that in both cases the two approaches used to impose constraint
preserving boundary conditions work very well in practice, with the
constraints converging to zero throughout the computational domain
even at late times, well after the initial pulses have reached the
boundaries. When comparing both approaches we find that the time
derivative method shows somewhat better convergence properties than
the space derivative method, particularly in the non-vacuum case.

As we mentioned previously, this approach to obtain boundary
conditions arises naturally from the hyperbolic structure of the
evolution system and may be generalized to the case of systems
including general matter fields coupled to gravity, and with less
restricting symmetry assumptions. However one must be cautious since
when considering cases whose domain has two or more effective spatial
dimensions there are more propagating eigenfields to be considered,
and also traverse derivatives of the eigenfields appear on the
analysis that need to be accounted for in a consistent way. Also, in
this work we have taken the shift vector to be an a priori known
function of space-time, but many state of the art simulations employ a
dynamical shift of the Gamma-driver family~\cite{Alcubierre02a}.  The
case of a Gamma-driver shift is conceptually similar to what we have
discussed here, but some important subtleties arise and we prefer to
leave that discussion for a further work.  Finally, even in the case
of axial symmetry, radiation (gravitational and otherwise) can not be
assumed to vanish so that for realistic scenarios one would need to
model the boundaries in a way that doesn't introduce spurious
radiation into the numerical domain (this is in fact a problem that
has been studied in many of the references given in the introduction,
but all the details have not yet been fully addressed~\cite{Friedrich:2009tq}).


\begin{acknowledgments}

This work was supported in part by CONACyT through grant 82787, by
DGAPA-UNAM through grants IN113907, IN115310 and DGAPA-PAPIIT
IN103514. J.~M.~T. also acknowledges a CONACyT postgraduate
scholarship.

\end{acknowledgments}


\appendix

\section{The constraint subsystem}
\label{sec:constraint.app}

Here we present an analysis of the characteristic structure of the
constraint subsystem. This analysis is based on that of
Ref.~\cite{Gundlach:2004jp}, with some modifications in order to adapt
it to our specific case. The constraints of the BSSN system satisfy by
themselves a homogeneous subsystem of evolution equations that
inherits its hyperbolicity properties from the full system. When
taking into account the extra constraint $C_\Delta$, and writing the
Ricci scalar that appears in the Hamiltonian constraint in terms of
the auxiliary variable $\Delta^r$, the constraint subsystem of
evolution in spherical symmetry reduces, up to principal part, to:
\begin{eqnarray}
\label{eq:constraint.subs}
\partial_0 H &\simeq& 0 \;, \\
\partial_0 M &\simeq&\frac{\alpha}{6} \partial_r H
+ \frac{\alpha}{2a}e^{-4\phi}\partial^2_r C_\Delta \;, \\
\partial_0 (\partial_r C_\Delta) &\simeq& 2\alpha\partial_r M \;.
\end{eqnarray}

\noindent One notices immediatly that the Hamiltonian constraint
behaves as a non-propagating eigenfield $\Omega_0=H$, while the two
combinations
\begin{equation}
\label{eq:constraint.Modes}
\Omega_\pm = \frac{a e^{4\phi} H}{6} + \frac{\partial_r C_\Delta}{2} 
\mp \sqrt{a}e^{2\phi} M \; ,
\end{equation}
propagate with eigenspeeds given by
\begin{equation}
  \label{eq:constraint.speeds}
  \lambda^\Omega_\pm =-\beta^r\pm\alpha e^{-2\phi}\sqrt{1/a} \; ,
\end{equation}
which again is are nothing more than the coordinate speed of light.

An important result regarding the constraint eigenfields $\Omega$ is
that each one is related to an eigenfield $\omega$ of the full
evolution system that propagates with the same speed, satisfying
\begin{equation*}
  \Omega \simeq s^i \partial_i \omega\;,
\end{equation*}
with $\vec{s}$ a vector that characterizes the propagation
direction. For our case it turns out that
$\lambda^l_\pm=\lambda^\Omega_\pm$ and thus
\begin{equation}
  \Omega_\pm \simeq  \partial_r \omega^l_\pm\;,
\end{equation}
which is the relation we use in order to construct constraint adapted
boundary conditions.


\bibliography{referencias}


\end{document}